\documentclass[aps,prb,twocolumn,superscriptaddress]{revtex4-1}

\usepackage{graphicx}
\usepackage{dcolumn}	
\usepackage{bm}			
\usepackage{here}
\usepackage{color}
\usepackage{ulem}
\begin{document}

\title{Localized-to-itinerant transition preceding antiferromagnetic quantum critical point and gapless superconductivity in CeRh$_{0.5}$Ir$_{0.5}$In$_5$}


\author{Shinji Kawasaki}
\thanks{To whom correspondence should be addressed; E-mail: kawasaki@science.okayama-u.ac.jp}
\affiliation{Department of Physics, Okayama University, Okayama 700-8530, Japan}

\author{Toshihide Oka}
\affiliation{Department of Physics, Okayama University, Okayama 700-8530, Japan}

\author{Akira Sorime}
\affiliation{Department of Physics, Okayama University, Okayama 700-8530, Japan}

\author{Yuji Kogame}
\affiliation{Department of Physics, Okayama University, Okayama 700-8530, Japan}

\author{Kazuhiro Uemoto}
\affiliation{Department of Physics, Okayama University, Okayama 700-8530, Japan}

\author{Kazuaki Matano}
\affiliation{Department of Physics, Okayama University, Okayama 700-8530, Japan}

\author{Jing Guo}
\affiliation{Institute of Physics, Chinese Academy of Sciences, and Beijing National Laboratory for Condensed Matter Physics,  Beijing 100190, China}

\author{Shu Cai}
\affiliation{Institute of Physics, Chinese Academy of Sciences, and Beijing National Laboratory for Condensed Matter Physics,  Beijing 100190, China}

\author{Liling Sun}
\affiliation{Institute of Physics, Chinese Academy of Sciences, and Beijing National Laboratory for Condensed Matter Physics,  Beijing 100190, China}

\author{John L. Sarrao}
\affiliation{Los Alamos National Laboratory, Los Alamos, New Mexico 87545, USA}

\author{Joe D. Thompson}
\affiliation{Los Alamos National Laboratory, Los Alamos, New Mexico 87545, USA}

\author{Guo-qing Zheng}
\thanks{To whom correspondence should be addressed; E-mail: gqzheng123@gmail.com}
\affiliation{Department of Physics, Okayama University, Okayama 700-8530, Japan}
\affiliation{Institute of Physics, Chinese Academy of Sciences, and Beijing National Laboratory for Condensed Matter Physics,  Beijing 100190, China}


\begin{abstract}
    
A fundamental problem posed from the study of correlated electron compounds, of which heavy-fermion systems are prototypes, is the need to understand the physics of states near a quantum critical point (QCP). At a QCP, magnetic order is suppressed continuously to zero temperature and unconventional superconductivity often appears. Here, we report pressure ($P$) -dependent $^{115}$In nuclear quadrupole resonance (NQR) measurements on heavy-fermion antiferromagnet CeRh$_{0.5}$Ir$_{0.5}$In$_5$.  These experiments reveal an antiferromagnetic (AF) QCP at $P_{\rm c}^{\rm AF}$ = 1.2  GPa where a dome of superconductivity reaches a maximum transition temperature $T_{\rm c}$. Preceding $P_{\rm c}^{\rm AF}$, however, the NQR frequency $\nu_{\rm Q}$ undergoes an abrupt increase at $P_{\rm c}^{\rm *}$ = 0.8 GPa in the zero-temperature limit, indicating a change from localized to itinerant character of cerium's $f$-electron and associated small-to-large change in the Fermi surface. At $P_{\rm c}^{\rm AF}$  where $T_{\rm c}$ is optimized, there is an unusually large fraction of gapless excitations well below $T_{\rm c}$ that implicates spin-singlet, odd-frequency pairing symmetry.

\end{abstract}


\maketitle


Understanding non-Fermi liquid behaviors \cite{Mathur,Lohneysen} due to a zero-temperature magnetic transition, a quantum critical point (QCP), and the unconventional superconductivity that emerges around it is one of the central issues in condensed matter physics. These phenomena are widely seen in strongly correlated electron systems such as heavy fermion systems \cite{Mathur}, cuprates \cite{Schmalian}, and iron pnictides \cite{Abraham,Zhou}. In cuprates, iron pnictides, or other compounds containing 3$d$ transition-metal elements \cite{JLuo}, the quantum phase transition is described by  itinerant spin-density wave (SDW) theories,  where the QCP is due to an instability of the underlying large Fermi surfaces \cite{Moriya_SCR,Hertz,Millis,SiSteglich}. Cerium(Ce)-based heavy fermion systems are understood  based on the Kondo lattice model in which  localized Ce 4$f$ electron spins at high temperatures are screened below a characteristic temperature $T_{\rm K}$ by the conduction electrons  \cite{Doniach}. At high temperatures, the $f$ electron spins are localized, and thus, the Fermi surface is  small. With decreasing  temperature, 4$f$ electrons couple with the conduction electrons through  Kondo hybridization, and the Fermi-surface volume gradually increases with decreasing temperature \cite{DLFengCo,DLFeng}. In addition to the Kondo effect, there also is a long-range Ruderman-Kittel-Kasuya-Yosida (RKKY) interaction that is the indirect exchange interaction among weakly screened, nearly localized $4f$ electrons. If the RKKY interaction overcomes the Kondo effect, $f$ spins order antiferromagnetically below the N\'{e}el temperature $T_{\rm N}$. By tuning a non-thermal control parameter such as pressure and/or chemical substitution,  $T_{\rm N}$ can be  suppressed completely to zero, and $T_{\rm K}$  increases as the parameter increases \cite{Doniach}. 
A crossover from the small (localized) to the large (itinerant) Fermi surfaces will occur well below $T_{\rm K}$ in the Kondo lattice \cite{Burdin}. Depending on the relative balance between Kondo hybridization and the RKKY interaction, magnetic order may be of the SDW-type or the RKKY-type antiferromagnetic (AF) order that is mediated by itinerant electrons of a small Fermi surface. Quantum criticality of the latter type of magnetic order is predicted theoretically to be qualitatively different from SDW criticality and involves a breakdown of Kondo screening and a transition from small-to-large Fermi surfaces at the QCP \cite{QSi, ColemanJLTP,SiSteglich}. In practice, it can be difficult experimentally to distinguish unambiguously between these two scenarios, though distinctions have been inferred from combinations of thermodynamic, transport and inelastic neutron scattering measurements \cite{LohneysenRMP}. Unfortunately, it has not been possible to perform neutron scattering experiments under high pressure conditions, the ``cleanest'' tuning parameter that does not introduce additional disorder or break symmetry, to shed light on the nature of criticality in Kondo lattice systems.  In contrast, pressure-dependent nuclear quadrupole resonance (NQR) measurements, which probe the dynamic spin susceptibility as well as the influence of Kondo hybridization, are straightforward, even at very low temperatures, and, as we show, can be used as differentiating probe of quantum criticality.

The antiferromagnetic superconductor CeRh$_{0.5}$Ir$_{0.5}$In$_5$ \cite{Pagliuso, Yamaguchi} is a good candidate to investigate this issue. CeRhIn$_5$ has a small Fermi surface \cite{Shishido}, orders antiferromagnetically below $T_{\rm N}$ = 3.8 K, and exhibits pressure-induced superconductivity above $P$ = 1.6 GPa at the transition temperature $T_{\rm c}$ = 2.1 K \cite{Hegger}. $T_{\rm c}$ increases to 2.3 K at 2.35 GPa where, in the limit of zero temperature, there is a magnetic-non-magnetic transition \cite{ParkNature,Knebel} accompanied by a change from small to large Fermi surface \cite{Shishido}. The superconducting-gap symmetry is consistent with $d$-wave  symmetry \cite{Mito2001,ParkPRL}. On the other hand,  CeIrIn$_5$ has a large Fermi surface \cite{ShishidoIr}, and also shows $d$-wave superconductivity below $T_{\rm c}$ = 0.4 K \cite{Petrovic1,ZhengIr,LuPRL}, which increases  to 0.8 K at $P$ = 2.1 GPa  \cite{KawasakiIr}. The alloyed compound CeRh$_{0.5}$Ir$_{0.5}$In$_5$ orders antiferromagnetically below $T_{\rm N}$ = 3.0 K at $P$ = 0 \cite{ADC} and becomes superconducting below $T_{\rm c}$ = 0.9 K \cite{Pagliuso,Nicklas,Yamaguchi}. CeRh$_{0.5}$Ir$_{0.5}$In$_5$ is closer to an antiferromagnetic QCP than CeRhIn$_5$, suggesting that Ir substitution for Rh acts a positive chemical pressure. This suggestion can be understood by appreciating that underlying the evolution of ground states in the CeRh$_{1-x}$Ir$_x$In$_5$ series is a systematic change in orbital anisotropy of the $\Gamma_7^2$ crystal-electric field ground state wavefunction that produces progressively stronger hybridization with increasing $x$ \cite{Willers}.

Here we report the results of $^{115}$In-NQR measurements on  CeRh$_{0.5}$Ir$_{0.5}$In$_5$ under pressure, crystal structure analysis and a first-principle calculation of NQR frequency $\nu_{\rm Q}$ (See Methods for details). From the temperature dependence of the nuclear spin-lattice relaxation rate ($1/T_1$), we find that the antiferromagnetic QCP is at $P_{\rm c}^{\rm AF}$ = 1.2 GPa, where $T_{\rm c}(P)$ reaches its maximum. From the pressure dependence of $\nu_{\rm Q}$, we find a localized-to-itinerant transition at  $P_{\rm c}^{\rm *}$ = 0.8 GPa before the antiferromagnetic QCP appears. Superconductivity is not only optimized at the antiferromagnetic QCP but also is realized with a remarkable proliferation of residual gapless excitations. Our results suggest that the large Fermi surface and antiferromagnetic instabilities in the presence of ``impurity'' scattering trigger unconventional gapless superconductivity in CeRh$_{0.5}$Ir$_{0.5}$In$_5$.

\begin{figure}
    \begin{center}
        \includegraphics[width=0.8\linewidth]{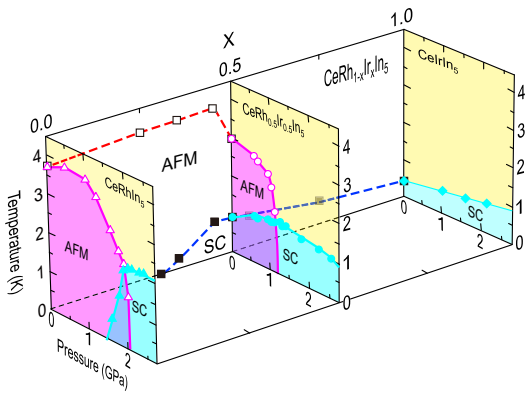}
    \end{center}
    \caption{{\bf Phase diagram of CeRh$_{1-x}$Ir$_x$In$_5$.} $x$ and pressure dependence of the N\'{e}el temperature $T_{\rm_N}$ (open squares, triangles, and circles) and the superconducting transition temperature $T_{\rm c}$ (solid squares, triangles, circles, and diamonds)  for  CeRh$_{1-x}$Ir$_x$In$_5$ ($P$ = 0) \cite{Yamaguchi}, CeRhIn$_5$ \cite{Hegger,ParkNature,Knebel}, and CeIrIn$_5$ \cite{KawasakiIr} under pressure.  AFM and SC indicate antiferromagnetic metal and superconductivity, respectively. }
    \label{f1}
\end{figure}

\section*{Results}

\noindent
\textbf{The hyperfine coupling constant at the In(1) site.}

The hyperfine coupling between nuclear and electronic spins relates the measured $1/T_1$ to the underlying dynamical spin susceptibility as discussed in Methods. Figure 1 shows the pressure-temperature phase diagram of the CeRh$_{1-x}$Ir$_{x}$In$_5$ system. If substituting Ir for Rh acts as a positive chemical pressure, we would expect the hyperfine coupling constant [$^{115}A(1)$] at the In(1) site (Fig. 2a) of CeRh$_{0.5}$Ir$_{0.5}$In$_5$ to be smaller than that of the host material CeRhIn$_5$ because $^{115}A(1)$ = 25 kOe $\mu_{\rm B}^{-1}$ in CeRhIn$_5$ decreases with increasing pressure but becomes constant at  $^{115}A(1)$ $\sim$ 7 kOe $\mu_{\rm B}^{-1}$ above $P$ = 1 GPa. Such a pressure dependent $^{115}A(1)$ will affect the information inferred from 1/$T_1$ \cite{Curro} and, therefore, it is important to determine $^{115}A(1)$ for CeRh$_{0.5}$Ir$_{0.5}$In$_5$ before proceeding to details of its $T_1$ results under pressure. Figure 2b shows the frequency-swept $^{115}$In-nuclear magnetic resonance (NMR) spectra at a constant field. The spectrum is consistent with previously reported spectra of CeRhIn$_5$ \cite{Curro}.   From these data, we establish the temperature dependence of the $^{115}$In(1)-NMR center line plotted in Fig. 2c. With these results, we calculate (Methods) the temperature dependence of the Knight shift $^{115}K(1)_{\rm c}$ (\%) and compare it to dc susceptibility $\chi_{\rm c}$ (emu mol$^{-1}$) in Fig. 2d. As clearly shown in the figure, the relation $K(T)$ $\propto$ $^{115}A(1)\chi(T)$ holds above $T$ = 30 K. A breakdown of this linear relationship at low temperature is common in heavy electron systems \cite{CurroYoung}, but a previous In-NMR study suggested that $^{115}A(1)$ is temperature-independent in CeIrIn$_5$ in such temperature regime \cite{Kambe1}. In the inset of Fig. 2d we plot $^{115}K(1)_{\rm c}$ versus $\chi_{\rm c}$ and obtain the hyperfine coupling constant as $^{115}A(1)$ = 7.64 kOe $\mu_{\rm B}^{-1}$. This closely corresponds to the value of $^{115}A(1)$ in CeRhIn$_5$ under a pressure of 1 GPa (Supplementary Note 1 and Supplementary Figure 1), and thus, it substantiates the suggestion that Ir substitution with $x$ = 0.5 (chemical pressure) is equivalent to the application of a  physical pressure greater than $P$ = 1 GPa to CeRhIn$_5$. Therefore, we reasonably can ignore any significant pressure dependence of  $^{115}A(1)$ in inferring the pressure evolution of physical properties from 1/$T_1$ in CeRh$_{0.5}$Ir$_{0.5}$In$_5$.

\begin{figure*}
    \centering
    \includegraphics[width=0.7\linewidth]{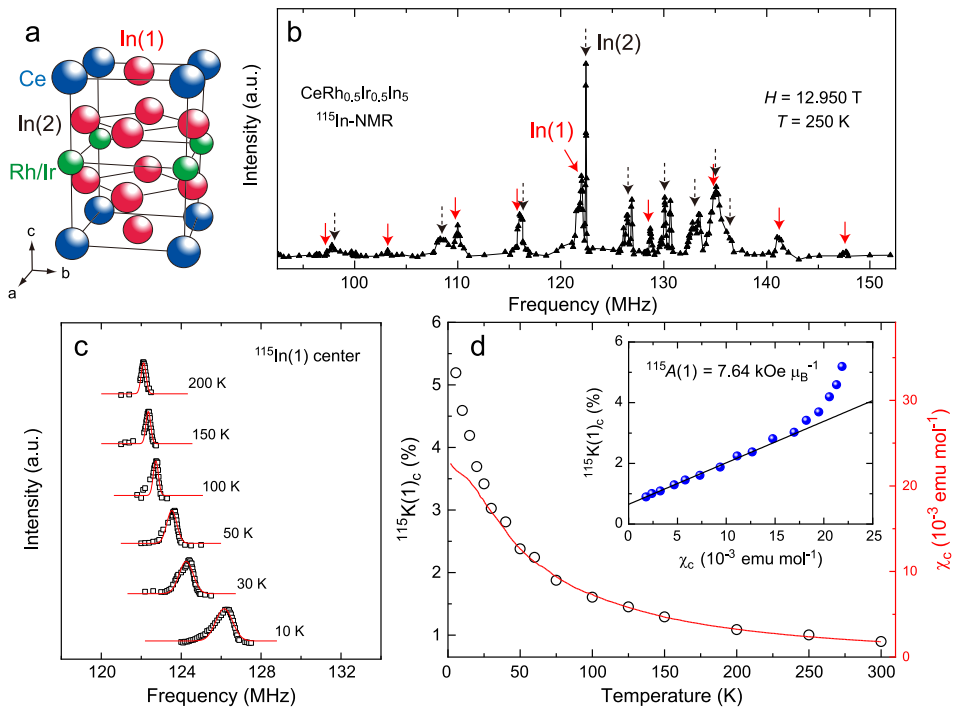}
    \caption{{\bf Hyperfine coupling constant.} (a) Crystal structure of CeRh$_{0.5}$Ir$_{0.5}$In$_5$. (b) Frequency-swept $^{115}$In-nuclear magnetic resonance (NMR) spectra of CeRh$_{0.5}$Ir$_{0.5}$In$_5$.  Solid [dotted] arrows indicate In(1)[In(2)] resonance peaks, respectively. (c) Temperature dependence of the $^{115}$In(1)-NMR center line.  The curves are Gaussian fits. (d) Temperature dependence of the Knight shift $^{115}$$K(1)_{\rm c}$ (open circles) and dc susceptibility  $\chi_{\rm c}$ (solid curve). The inset shows $^{115}$$K(1)_{\rm c}$ versus $\chi_{\rm c}$. The solid straight line is a fit to the data above $T$ = 30 K which yields the hyperfine coupling constant $^{115}$$A(1)$ = 7.64 kOe $\mu_{\rm B}^{-1}$. Error bars are smaller than the size of the data points. }  
\end{figure*}

\begin{figure}
    \begin{center}
        \includegraphics[width=0.85\linewidth]{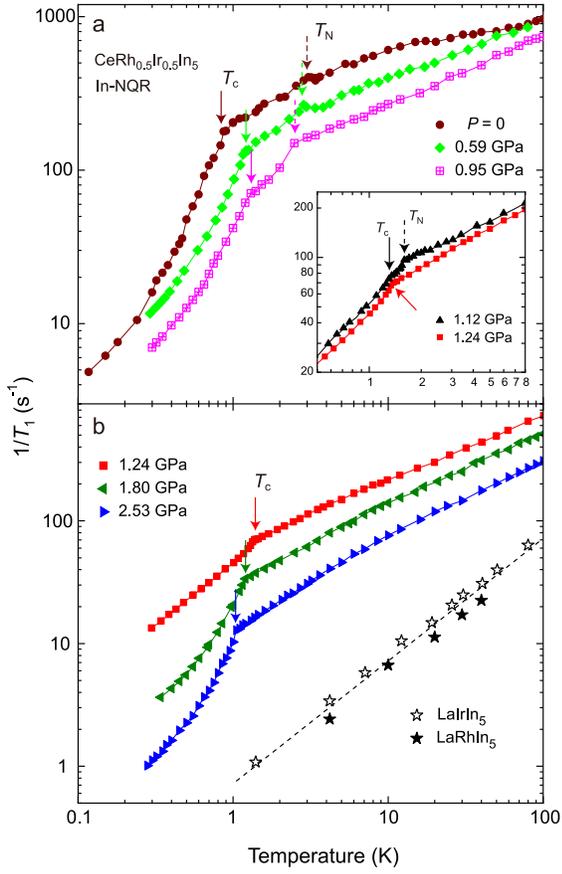}
    \end{center}
    \caption{{\bf Antiferromagnetism and superconductivity under pressure.} Temperature dependence of the In(1)-nuclear quadrupole resonance (NQR) nuclear spin-lattice relaxation rate 1/$T_1$ (a) below and (b) above the antiferromagnetic (AF) quantum critical point  $P_{\rm c}^{\rm AF}$.  (Inset) Data for $P$ = 1.12 and 1.24 GPa just around the N\'{e}el temperature $T_{\rm N}$ and the superconducting transition temperature $T_{\rm c}$. The dotted (solid) arrows indicate $T_{\rm N} (T_{\rm c})$, while the dashed line for La(Ir,Rh)In$_5$ indicates 1/$T_1T$ = constant. Data at $P$ = 0 \cite{Yamaguchi} and for La(Ir,Rh)In$_5$ are obtained from \cite{GQZ}. Error bars are smaller than the size of the data points.} 
\end{figure}

\begin{figure*}
    \begin{center}
        \includegraphics[width=0.98\linewidth]{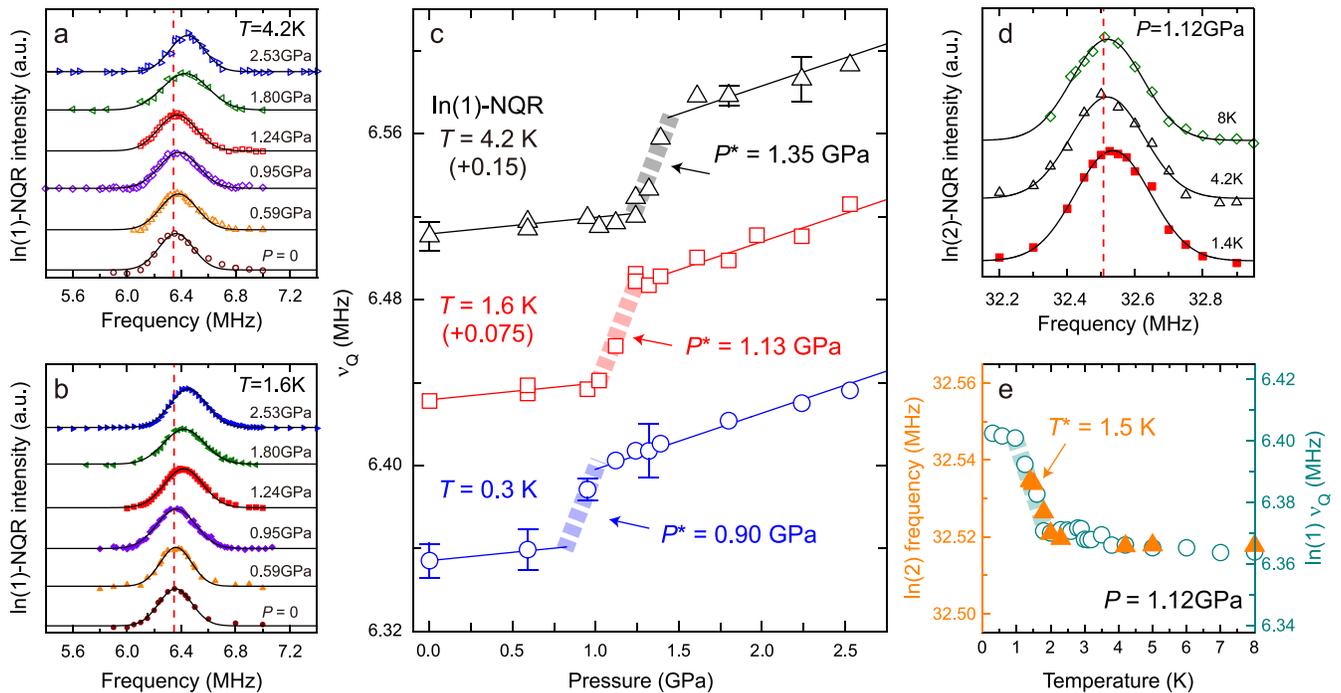}
    \end{center}
    \caption{{\bf Localized to itinerant transition.} Pressure dependence of the In(1) $^{115}$In-nuclear quadrupole resonance (NQR) spectrum 
        ($\pm$1/2 $\leftrightarrow$ $\pm$3/2 transition line) at $T$ = 4.2 K (a) and 1.6 K (b).  The curves are Gaussian fits.  The dotted vertical lines indicate the peak positions at $P$ = 0. (c) Pressure dependence of the NQR frequency ($\nu_{\rm Q}$) obtained at $T$ = 0.3, 1.6, and 4.2 K.  For clarity, the vertical axis for $T$ = 1.6 K and 4.2 K are offset by the amount shown in the parenthesis. Solid arrows indicate localized-to-itinerant transition pressure ($P^{\rm *}$) which is defined from the midpoint of $\nu_{\rm Q}$ jump.  Solid straight lines are fits to the data which yield the slope $d\nu_{\rm Q}/dP$ = 0.008 (0.027) below (above) $P^{\rm *}$. Error bars represent the uncertainty in estimating $\nu_{\rm Q}$.  (d) Temperature dependence of the In(2) $^{115}$In-NQR spectrum  ($\pm$3/2 $\leftrightarrow$ $\pm$5/2 transition line) at $P$ = 1.12 GPa. The curves are Gaussian fits. Vertical line indicates the peak position at $T$ = 8 K. (e) Temperature  dependence of the peak position of the In(2) spectrum (solid triangles) and In(1) $\nu_{\rm Q}$ (open circles) at $P$ = 1.12 GPa.  Solid arrow indicates localized-to-itinerant transition temperature $T^{\rm *}$ which is defined from the midpoint of $\nu_{\rm Q}$ jump. The dashed line serves as visual guide. Error bars are smaller than the size of the data points except for those in (c).} 
\end{figure*}

\begin{figure}
    \begin{center}
        \includegraphics[width=0.9\linewidth]{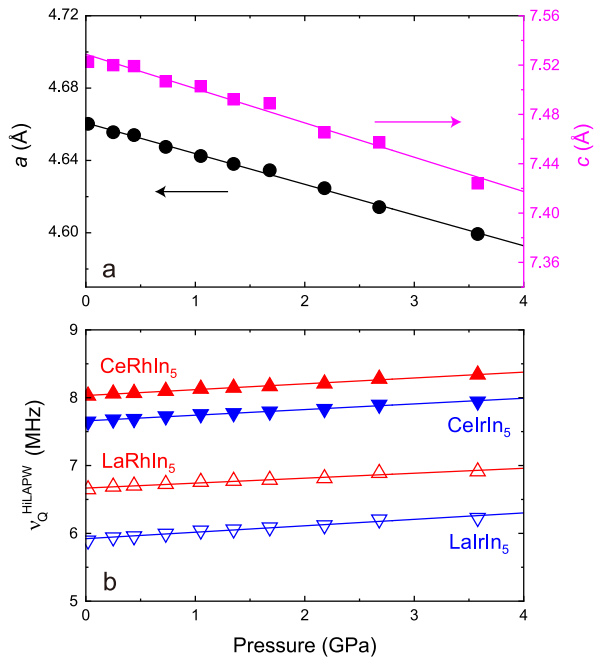}
    \end{center}
    \caption{{\bf Lattice parameters and band calculation.} (a) Pressure dependence of $a$- and $c$-axis lengths for CeRh$_{0.5}$Ir$_{0.5}$In$_5$ at room temperature. (b) Pressure dependence of the nuclear quardrupole resonance frequency $\nu_{\rm Q}^{\rm HiLAPW}$ obtained from band calculations [Hiroshima Linear-Augmented-Plane-Wave (HiLAPW) code] for CeRh(Ir)In$_5$ and LaRh(Ir)In$_5$. LaRh(Ir)In$_5$ corresponds to the 4$f$-localized model of CeRh(Ir)In$_5$. Both are calculated with the same lattice constant. Error bars are smaller than the size of the data points.} 
\end{figure}

\begin{figure*}
    \centering
    \includegraphics[width=0.6\linewidth]{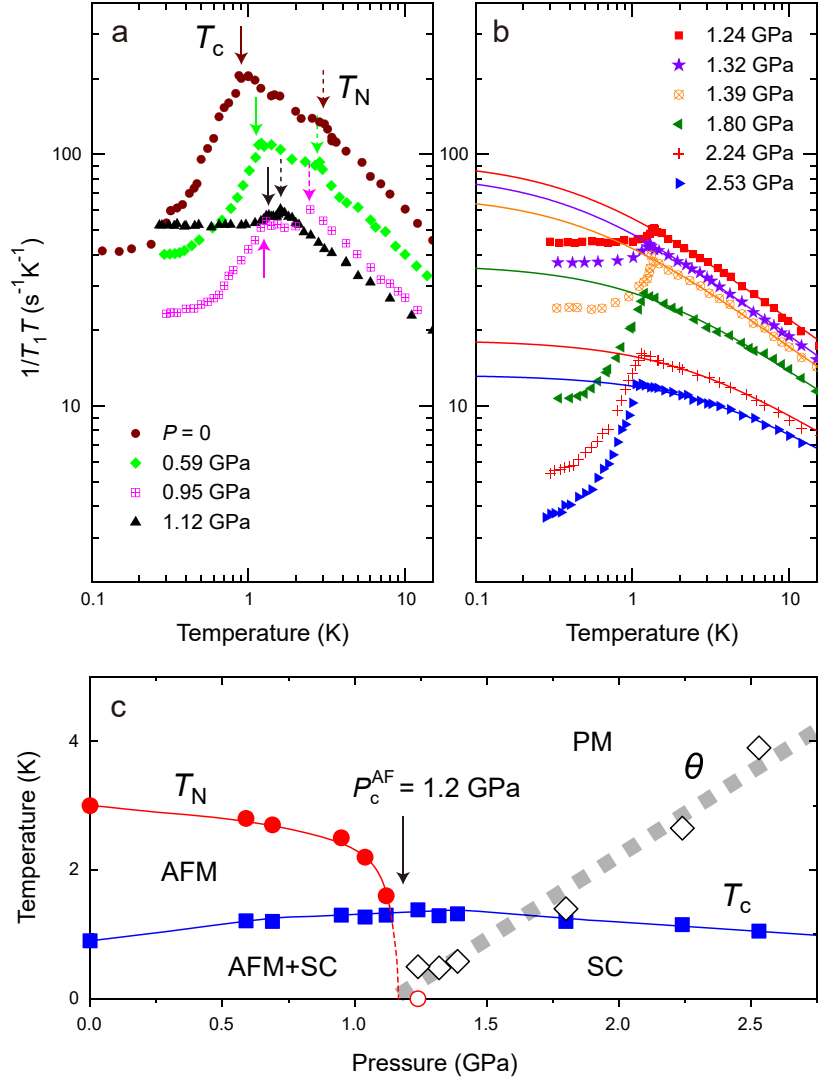}
    \caption{{\bf Antiferromagnetic (AF) quantum critical point ($P_{\rm c}^{\rm AF}$).} Temperature dependence of the nuclear spin-lattice relaxation rate divided by temperature $1/T_1T$ below (a) and above (b) $P_{\rm c}^{\rm AF}$ with the 
        fitting curves (see text).  The dotted (solid) arrows indicate the N\'{e}el temperature $T_{\rm N}$ (the superconducting transition temperature $T_{\rm c})$. (c) Summary of pressure dependence of the characteristic temperature $\theta$ (see text), $T_{\rm N}$, and $T_{\rm c}$. The dotted straight line (filled diamonds) is a fit  to the data which yields 
        $P_{\rm c}^{\rm AF} $ ($\theta = 0$) = 1.2 GPa. PM and AFM indicate paramagnetic and antiferromagnetic metal, respectively. SC indicates superconductivity. Error bars are smaller than the size of the data points.    } 
\end{figure*}

\begin{figure*}
    \centering
    \includegraphics[width=0.8\linewidth]{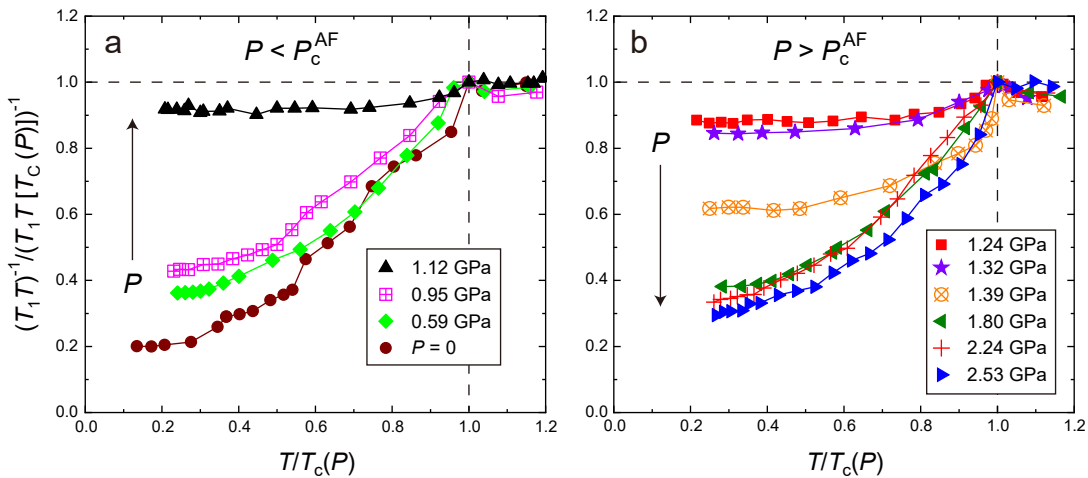}
    \caption{{\bf Low-lying excitations in the superconducting state.} Temperature dependence of the nuclear spin-lattice relaxation rate divided by temperature 1/$T_1T$ below (a) and above (b) the antiferromagnetic (AF) quantum critical point ($P_{\rm c}^{\rm AF}$).  Data at $P$ = 0 are obtained from the literature \cite{Yamaguchi}.  Horizontal and vertical axes are normalized by the superconducting transition temperature $T_{\rm c}(P)$ and $1/T_1T[T_{\rm c}(P)]$, respectively. Error bars are smaller than the size of the data points.}
    
\end{figure*}

\begin{figure*}
    \centering
    \includegraphics[width=0.65\linewidth]{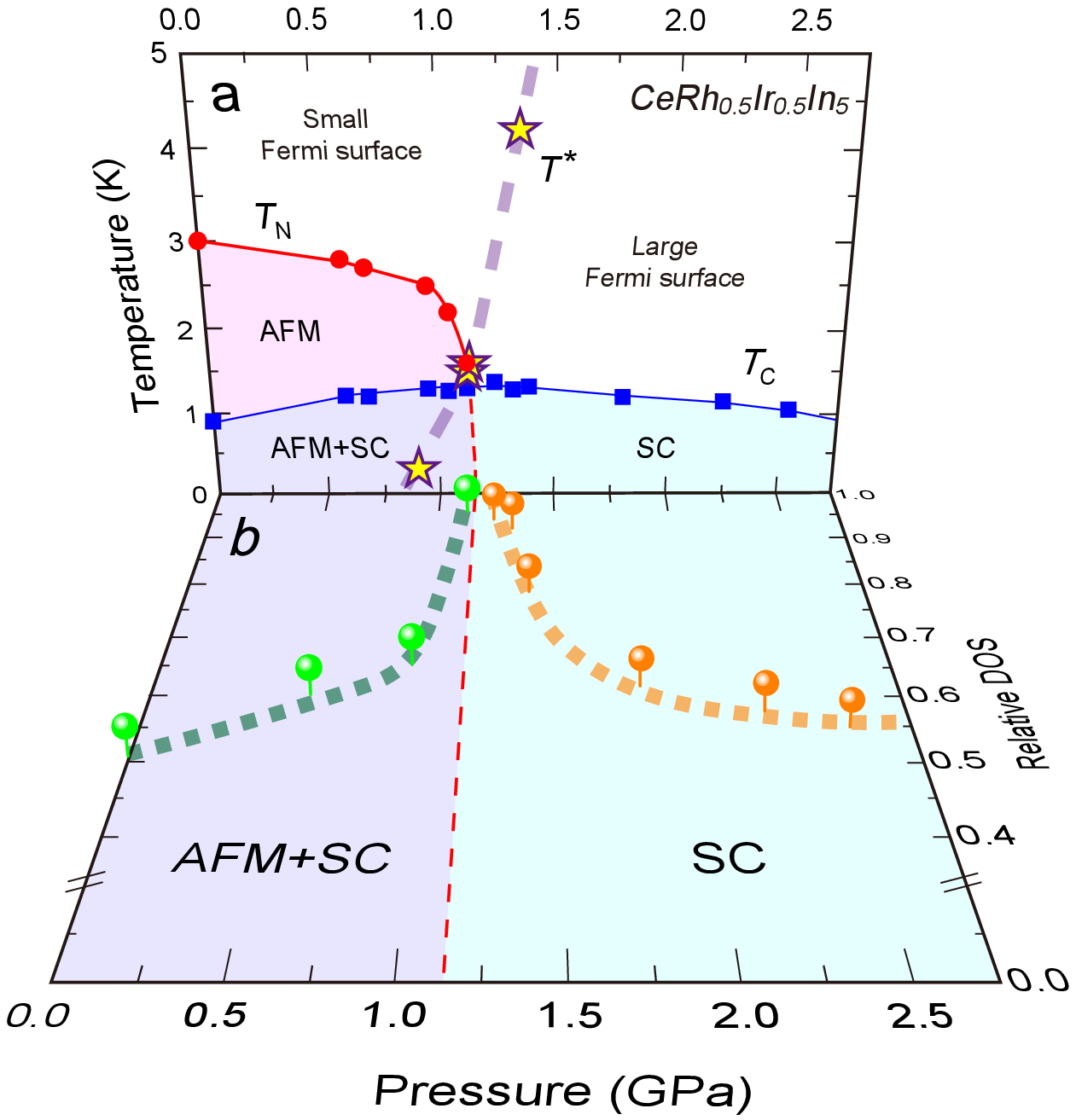}
    \caption{{\bf Phase diagram.} (a) Pressure-temperature phase diagram for  CeRh$_{0.5}$Ir$_{0.5}$In$_5$ obtained at zero-magnetic field. The solid circles, squares, and stars indicate the N\'{e}el temperature $T_{\rm N}$, the superconducting transition temperature $T_{\rm c}$, and the localized-to-itinerant transition temperature $T^{*}$, respectively. The broken curve is a phase boundary separating small and large Fermi surfaces.  (b) Pressure dependence of the relative density of states (DOS) on the $T$ = 0.3 K plane.  The dotted straight line indicates the antiferromagnetic (AF) quantum critical point ($P_{\rm c}^{\rm AF}$ = 1.2 GPa). The solid and dashed curves serve as visual guides. AFM and SC indicate antiferromagnetic metal and superconductivity, respectively.  Error bars are smaller than the size of the data points.}  
\end{figure*}

\vspace{1cm}
\noindent
\textbf{Pressure dependence of $T_{\rm N}$ and $T_{\rm c}$.}
Figure 3a and 3b show the temperature dependence of $1/T_1$.  The magnitude of $1/T_1 $ for CeRh$_{0.5}$Ir$_{0.5}$In$_5$ is much greater than that for the nonmagnetic reference material La(Rh,Ir)In$_5$ \cite{GQZ}, due to $f$ electron spin correlations, that are reflected in the dynamical susceptibility $\chi ^{\prime\prime}$ to which $1/T_1$ is proportional. At $P$ = 0, $1/T_1 $ exhibits a small peak at $T_{\rm N}$ = 3.0 K and decreases below $T_{\rm c}$ = 0.9 K \cite{Yamaguchi}. As shown in Fig. 3a and the inset, $T_{\rm N}(P)$ can be identified up to $P$ = 1.12 GPa but disappears above $P_{\rm c}^{\rm AF}$ = 1.2 GPa. A superconducting transition is observed at all pressures, as evidenced by an abrupt reduction of 1/$T_1$ below $T$ = $T_{\rm c}(P)$.  $T_{\rm c}$ increases with increasing pressure and exhibits a maximum $T_{\rm c}^{\rm max}$ = 1.4 K around $P_{\rm c}^{\rm AF}$ = 1.2 GPa and then decreases with further increase of $P$; $T_{\rm c}^{\rm max}$ is 1.6 times higher than $T_{\rm c}$ = 0.9 K at $P$ = 0.

\vspace{1cm}
\noindent
\textbf{Pressure dependence of the Kondo hybridization.}
To probe the character of $f$ electrons as a function of pressure, we use the In(1) $^{115}$In-NQR frequency $\nu_{\rm Q}$. In general, $\nu_{\rm Q}$ is determined by the surrounding lattice and on-site electrons with the latter being dominant in strongly correlated electron systems \cite{ZhengJPSJ}; in the current case the latter reflects $f$-$c$ hybridization that generates an electric field gradient (EFG)  at the In(1)-site, as was found in previous $^{115}$In-NQR and NMR studies on CeIn$_3$ \cite{KawasakiCeIn3} and CeRhIn$_5$ \cite{Curro} under pressure. Figures 4a, 4b, and 4c show the pressure dependence of the In(1) NQR spectrum and $\nu_{\rm Q}$ (see Supplementary Note 2 and Supplementary Figure 2 for detail). In general, $\nu_{\rm Q}$ is expected to increase smoothly with decreasing volume \cite{KawasakiCeIn3}; however, this is not the case here. At $T$ = 4.2 K, $\nu_{\rm Q}$ weakly increases up to $P$ = 1.24 GPa but jumps at $P^{\rm *}$ = 1.35 GPa above which the slope $d\nu_{\rm Q}/dP$  increases by more than a factor of three. The same trend is found at lower temperatures where we see that $P^{\rm *}$ decreases as $T$ is reduced.  We denote the midpoint of the $\nu_{\rm Q}$ jump in the $P$-$T$ plane  as ($P^{\rm *}$, $T^{\rm *}$) = (1.35 GPa, 4.2 K), (1.13 GPa, 1.6 K), and (0.90 GPa, 0.3 K), respectively.  The same result is also found in the In(2) site. Figures 4d and 4e show the temperature dependence of the In(2) NQR spectrum and its peak position together with In(1) $\nu_{\rm Q}$ at $P$ = 1.12 GPa. As found in the In(1) site, the In(2) $\nu_{\rm Q}$ increases below $T^*$ = 1.5 K.  In a previous study on CeRhIn$_5$, a similar change of EFG was found in the In(2) site at $P^{\rm *}$ = 1.75 GPa, although detailed pressure dependence  for the In(1) site with a high accuracy was not  conducted \cite{Curro}.

We consider the origin for the $\nu_{\rm Q}$ jump. Figure 5a shows results of X-ray diffraction measurements  that  give  the pressure dependence of $a$- and $c$-axis lengths; the lattice parameters decrease smoothly with increasing pressure to $P$ = 4 GPa, without any signature of a structural transition.  From these data, we calculate the EFG using the first-principles Hiroshima Linear-Augmented-Plane-Wave (HiLAPW) codes \cite{Oguchi}. 
As expected,  the calculated $\nu_{\rm Q}^{\rm HiLAPW}$ increases monotonically with applied pressure (Fig. 5b). This and the lack of any anomaly in 1/$T_1$ at ($P^{\rm *}, T^{\rm *}$) rule out a change in lattice contribution to $\nu_{\rm Q}$ as the origin of the $\nu_{\rm Q}$  jump.  Furthermore, Fig. 5b compares the pressure dependence of $\nu_{\rm Q}^{\rm HiLAPW}$ for CeRh(Ir)In$_5$ and LaRh(Ir)In$_5$. $\nu_{\rm Q}^{\rm HiLAPW}$ for CeRh(Ir)In$_5$ is uniformly greater than that for LaRh(Ir)In$_5$, because CeRh(Ir)In$_5$ has the additional EFG from hybridized $f$ electrons, unlike non-magnetic LaRh(Ir)In$_5$. This result is consistent with previous band calculations \cite{Harima,CurroBandCalc}.

We conclude from these results that the jump in $\nu_{\rm Q}(P)$ at ($P^{\rm *}$, $T^{\rm *}$) is due to a pronounced increase in Kondo hybridization at ($P^{\rm *}$, $T^{\rm *}$) and that the larger $d\nu_{\rm Q}/dP$ above ($P^{\rm *}$, $T^{\rm *}$) reflects that increased hybridization. Because increased Kondo hybridization transfers $f$ spectral weight from localized to itinerant degrees of freedom, and hence an increase in Fermi surface volume, the pronounced jump in $\nu_{\rm Q}$ signals the experimental observation of a small (localized) to large (itinerant) Fermi surface. As the principal axis of the EFG at the In(2) site is perpendicular to that of the In(1) site, the $\nu_{\rm Q}$ change at both sites suggests that the entire Fermi-surface volume changes at ($P^*$, $T^*$).

  \vspace{1cm}
 \noindent
  \textbf{Determination of the antiferromagnetic QCP.}
  The temperature dependence of $T_1$ just above $T$ $\geq$ $T_{\rm c}(P)$ can be described by the self-consistent renormalization (SCR) theory for spin fluctuations around an antiferromagnetic QCP \cite{Moriya}. A three-dimensional antiferromagnetic spin fluctuation model is applicable also to the low temperature thermopower $S/T$ around the pressure-induced antiferromagnetic QCP of CeRh$_{0.58}$Ir$_{0.42}$In$_5$ \cite{Luo}.  Near an antiferromagnetic QCP,  the SCR model predicts $1/T_1T \propto \sqrt{\chi_{\rm {\bf Q}} (T)} = 1/\sqrt{T+\theta}$ \cite{Moriya} where  $\chi_{\rm {\bf Q}} (T)$ is the Curie-Weiss staggered susceptibility and $\theta$ is a measure of the distance to the antiferromagnetic QCP. At the QCP, $\theta = 0$ and $\chi_{\rm {\bf Q}} (T)$ diverges toward 0 K. $1/T_1T$ can be represented by the sum of magnetic and small non-magnetic contributions as $1/T_1T$ = $1/(T_1T)_{\rm AF} + 1/(T_1T)_{\rm lattice}$.  For the lattice term, we use the mean value of $1/T_1T$ from reference materials LaRhIn$_5$ and LaIrIn$_5$ (Fig. 3b), which gives $1/(T_1T)_{\rm lattice}$ = 1.44 (s$^{-1}$K$^{-1}$).

Figures 6a and 6b are plots of $1/T_1T$ vs. $T$ for the antiferromagnetic phase below $P$ = 1.12 GPa and for the paramagnetic phase above $P$ = 1.24 GPa, respectively.  
The solid curves in Fig. 6b are least-squares fits to 1/$T_1T$ = 
$a/(T+\theta)^{0.5}$ + 1.44 just above $T_{\rm c}(P)$, with $a$ and $\theta$ as  parameters. Approaching the antiferromagnetic QCP from $P$ = 2.53 GPa, $\theta$ decreases with decreasing pressure, as can be seen in Fig. 6b. The pressure dependences of $\theta$,  $T_{\rm N}$, and $T_{\rm c}$ are plotted in Fig. 6c. From a linear fit of $\theta(P)$, the antiferromagnetic QCP ($\theta$ = 0 K) for CeRh$_{0.5}$Ir$_{0.5}$In$_5$ is obtained at 
$P_{\rm c}^{\rm AF}$ = 1.2 GPa, where the highest $T_{\rm c}$ is realized. The present results clearly indicate that spin fluctuations play a significant role for superconductivity in CeRh$_{0.5}$Ir$_{0.5}$In$_5$.

\vspace{1cm}
\noindent
\textbf{Unconventional superconductivity at $P_{\rm c}^{\rm AF}$.} 
As seen in Figs. 6a and 6b, there is a strong pressure dependence of the magnitude of 1/$T_1T$ at the lowest temperatures of these measurements. To place these results in perspective, we normalize 1/$T_1T$ by its value at $T_{\rm c}$, 1/$T_1T(T_{\rm c})$, and plot the ratio in Figs. 7a and 7b as a function of reduced temperature $T/T_{\rm c}(P)$. Deep in the superconducting state, this ratio is clearly largest at $P$ = 1.12 GPa near $P_{\rm c}^{\rm AF}$ and depends only weakly on temperature below $T_{\rm c}$. This result contrasts with expectations for a fully gapped, e.g., $s$-wave, superconductor where 1/$T_1$ should decrease exponentially to a very small value well below $T_{\rm c}$ and for a clean $d$-wave superconductor where 1/$T_1$ decreases as $T^3$. In CeRh$_{0.5}$Ir$_{0.5}$In$_5$ at $P$ = 1.12 GPa there must be a substantial fraction of low-lying excitations in the normal state that remains ungapped below $T_{\rm c}$. Namely,  [$T_1T$($T$=0.3K)$]^{-0.5}$/[$T_1T(T_{\rm c})]^{-0.5}$ = $N(E_{\rm F})^{\rm residual}$/$N(E_{\rm F})^{\rm normal}$ is the relative density of state (DOS) at  $T$ = 0.3 K, which is consistent with the fraction of ungapped quasiparticle DOS in the superconducting state. 

To obtain the relative DOS, for simplicity we assume that $T_1$  below $T_{\rm c}$ is predominantly determined by itinerant quasi particles. Previously, an analysis within the context of two-fluid phenomenological theory has deduced that the  4$f$ local moments  also contribute to relaxation \cite{PinesPNAS,Curro2fluid}. Nonetheless, as shown in Supplementary Note 3 and Supplementary Figures 3 and 4, such a model reproduces  essentially the same result as we obtained here.

\vspace{1cm}
\noindent
\textbf{Phase Diagram.} 
Figure 8a shows the pressure dependence of $T_{\rm N}$ and $T^{\rm *}$ inferred from In(1) NQR, together with $T_{\rm c}$ on the $P$-$T$ plane. $T^{\rm *}$ inferred from In(2) NQR at $P$ = 1.12 GPa coincides with the result obtained from In(1). $T^{\rm *}$  extrapolates to zero at $P_{\rm c}^{\rm *}$ = 0.8 GPa, which is distinctly smaller than $P_{\rm c}^{\rm AF}$ = 1.2 GPa. In CeRhIn$_5$ ($x$ = 0), a similar result was suggested with $P_{\rm c}^{\rm *}$ = 1.75 GPa \cite{Curro} and $P_{\rm c}^{\rm AF}$ = 2.1 GPa \cite{Yashima2007}.  Comparison of these critical pressure values shows that  Ir substitution with $x$ = 0.5 (chemical pressure) is effectively equivalent to the application of a  physical pressure of about $P$ = 1 GPa to CeRhIn$_5$.  This conclusion is consistent with that drawn from our measurement of the  hyperfine coupling presented earlier (Supplementary Note 1 and Supplementary Figure 1). We emphasize again that, in the pressure regions we are interested for CeRhIn$_5$ and CeRh$_{0.5}$Ir$_{0.5}$In$_5$, $^{115}A(1)$  is constant, and thus changes in physical properties are not related to a changing hyperfine coupling but to the quantum criticality.

As shown in Fig. 8b, in CeRh$_{0.5}$Ir$_{0.5}$In$_5$, remarkably, the fraction of ungapped excitations strongly depends on pressure, reaching a maximum at the antiferromagnetic QCP. In the coexistent state at $P$ = 0 \cite{Yamaguchi}, this fraction is 50 \%, but increases to 96\% at $P_{\rm c}^{\rm AF}$ and then decreases to 55\% with the increasing pressure at $P$ = 2.53 GPa. The highest $T_{\rm c}$ around the antiferromagnetic QCP of CeRh$_{0.5}$Ir$_{0.5}$In$_5$ is realized unexpectedly with the largest fraction of gapless excitations. The present observation is completely different from that for CeRhIn$_5$ under pressure; in CeRhIn$_5$, the relative fraction of gapless excitations (88\%) in the coexistent state is rapidly suppressed to almost zero as it approaches the QCP \cite{Yashima2007}.

\section{Discussion}

From the phase diagram shown in Fig. 8, it is clear that the localized to itinerant transition ($T^{\rm *}(P)$) does not occur exactly at the antiferromagnetic QCP; in the limit of zero temperature, $P_{\rm c}^{\rm *}$ precedes $P_{\rm c}^{\rm AF}$. One possibility would be that the Fermi-surface change across the $T^{\rm *}(P)$ boundary marks a line of abrupt changes of the Ce valence that terminates near $T$ = 0 in a critical end point. A model that considers this possibility, however, appears to exclude the $T^{\rm *}(P)$ boundary from extending into the antiferromagnetic state \cite{MiyakeWatanabe,MiyakeWatanabe2014}, contrary to our results.  In contrast, a breakdown of the Kondo effect gives rise to a small to large Fermi surface change across $T^{\rm *}(P)$ \cite{QSiT1}. This idea leads to a $T$ = 0 phase diagram \cite{QSiPhysB,ColemanJLTP} similar to the results of Fig. 8. Associated with the Kondo breakdown and development of a large Fermi surface, soft charge fluctuations can emerge without a change in formal valence of Ce ions \cite{Komijani}. Within experimental uncertainty of $\pm$1.5 \%, there is no detectable difference between CeRhIn$_5$ and CeIrIn$_5$ at 10 K in their spectroscopically determined Ce valence \cite{Sundermann}, even though their Fermi volumes differ - a result that, together with the phase diagram of Fig. 8, supports a Kondo breakdown interpretation as do thermopower measurements on CeRh$_{0.58}$Ir$_{0.42}$In$_5$ \cite{Luo}.  Notably, the $T^{\rm *}$ boundary has no notable effect on the evolution of $T_{\rm c}(P)$. In passing, we mention a possibility of a more general case, a Lifshitz transition. Watanabe and Ogata \cite{Ogata}, and Kuramoto $et$ $al$ \cite{Kuramoto} pointed out that, even though the Kondo screening remains,  a competition between the Kondo effect and the RKKY interaction can lead to a topological Fermi surface transition (Lifshitz transition)  below the antiferromagnetic QCP \cite{Ogata,Kuramoto}, which is also consistent with our observation.

The large $1/T_1T$ below $T_{\rm c}$ at pressures near the antiferromagnetic QCP (Figs. 6 and 7) is quite remarkable, and its temperature independence implies a large DOS that remains ungapped in the superconducting state even though this is the pressure range where $T_{\rm c}$ is a maximum. Such an anomalously high value of ungapped DOS  in the superconducting state has never been found in other QCP materials such as high $T_{\rm c}$ cuprates \cite{Asayama} and the iron-pnictide superconductors \cite{Zhou,Oka}, even though magnetic fluctuations are also strong around their QCP. For the Ce115 family, no significant gapless excitations have been observed so far around a QCP \cite{Kohori}.

 In general, gapless excitations are expected from impurity scattering in $d$-wave superconductors \cite{MiyakeVarma,Hirschfeld,Maebashi,Haas}. Though there are no intentionally added impurities in our crystal, the random replacement of 50 \% Rh by Ir results in a broadening of the In(1) NQR line by a factor of  $\sim$ 5 compared to CeIrIn$_5$ \cite{Yamaguchi,ZhengIr}. Such randomness increases the resistivity at 4 K (just above $T_{\rm N}$) from $\sim$ 4 $\mu\Omega$cm in CeRhIn$_5$ to over 20 $\mu\Omega$cm in CeRh$_{0.5}$Ir$_{0.5}$In$_5$ \cite{Nicklas}. Quantum critical fluctuations can further enhance that scattering \cite{Maebashi} to make part of a multi-sheeted Fermi surface gapless  \cite{Barzykin}.
Such scattering concomitantly leads to pair breaking, resulting in  a large reduction of $T_{\rm c}$ \cite{Abrikosov,MiyakeVarma,Hirschfeld,Maebashi,Haas,Barzykin}, which is inconsistent with our observations. 
The relative DOS  at the pressure-induced QCP is almost zero in CeRhIn$_5$ \cite{Yashima2007} but is enhanced  to 96\% in CeRh$_{0.5}$Ir$_{0.5}$In$_5$ at $P_{\rm c}^{\rm AF}$. If we assume that the symmetry of CeRh$_{0.5}$Ir$_{0.5}$In$_5$ is also $d$-wave, $T_{\rm c}$ should be reduced to zero with such a significant residual DOS at $E_{\rm F}$ \cite{Maki}. In contrast to this expectation, the maximum  $T_{\rm c}$ = 1.4 K for CeRh$_{0.5}$Ir$_{0.5}$In$_5$ remains at 61\% of $T_{\rm 	c}$ = 2.3 K for CeRhIn$_5$ at their respective QCPs.   
Hence, the present results suggest that superconductivity near $P_{\rm c}^{\rm AF}$ in CeRh$_{0.5}$Ir$_{0.5}$In$_5$ is more exotic than $d$-wave.

Model calculations of superconductivity in a two-dimensional Kondo lattice show that near an antiferromagnetic QCP $d$- and $p$-wave spin-singlet superconducting states are nearly degenerate, with an odd frequency $p$-wave spin-singlet state being favored when entering the  large Fermi surfaces  region to take advantage of the nesting condition with the vector ${\bf Q}$ = ($\pi$,$\pi$) \cite{Otsuki}.  A $p$-orbital wave function with spin-singlet pairing symmetry satisfies Fermi statistics in the odd-frequency channel \cite{Berezinskii,Coleman,Balatsky,Fuseya}, and this odd-frequency pairing is more robust against non-magnetic impurity scattering than even-frequency pairing \cite{Yoshioka}. Indeed, for a scattering strength that kills $d$-wave superconductivity completely, spin-singlet odd-frequency pairing will survive with a $T_{\rm c}$ that is approximately 60 \% of that in the absence of scattering \cite{Yoshioka}. Motivated by these theoretical results, we suggest that odd-frequency spin-singlet pairing is realized in CeRh$_{\rm 0.5}$Ir$_{0.5}$In$_5$ in the vicinity of its critical pressures. Its robust $T_{\rm c}$ in the presence of substantial disorder scattering that gives rise to a large residual density of states at $P_{\rm c}^{\rm AF}$ where quantum critical fluctuations are strongest and the presence of a nearby change from small to large Fermi surface at $P_{\rm c}^{*}$ are fully consistent with our proposal. We stress that the unique aspect of both strong fluctuations and large Fermi surface is not shared by the end members, CeIrIn$_5$ or CeRhIn$_5$. Knight shift and experiments that directly probe the gap symmetry will be useful to test this possibility.

 In summary, we reported systematic $^{115}$In-NQR measurements on the heavy fermion 
 antiferromagnetic superconductor CeRh$_{0.5}$Ir$_{0.5}$In$_5$ under pressure and find that an antiferromagnetic QCP is located at $P_{\rm c}^{\rm AF}$ = 1.2 GPa, at which $T_{\rm c}(P)$ reaches its maximum. The pressure and temperature dependence of $\nu_{\rm Q}$ reveal a pronounced increase in hybridization that signals a change from small to large Fermi surface in the limit of zero temperature at $P_{\rm c}^{\rm *}$ = 0.8 GPa which is notably lower than $P_{\rm c}^{\rm AF}$. Thus, our work sheds new light on the quantum phase transition in $f$-electron systems. There is a strong enhancement of the quasiparticle DOS in the superconducting state around $P_{\rm c}^{\rm AF}$ where the Fermi surface is large. The robustness of $T_{\rm c}$ under these conditions can be understood if the superconductivity is odd-frequency $p$-wave spin singlet. Traditionally, Hall coefficient and quantum oscillation experiments have been used to probe the Fermi surface change.
 Our work demonstrates that the NQR frequency can be used as a powerful tool to examine the change in Fermi surface volume for heavy electron systems. In particular, the NQR technique does not require single crystals and  is not limited by sample quality or pressure, and thus will open a new venue  to understand strongly correlated electron superconductivity.

 \section*{Methods}
\noindent
\textbf{Samples.}

 Single crystals of CeRh$_{0.5}$Ir$_{0.5}$In$_5$ were grown from an In flux as reported in a previous study \cite{Pagliuso}. All experiments were performed with the same batch of crystals used in the previous NQR study \cite{Yamaguchi}. As documented in detail in Ref. \cite{Yamaguchi}, there is no phase separation into Rh-rich and Ir-rich parts even in the coarsely crushed powder. In fact, no excess peaks in the NMR/NQR spectrum are found and the spectrum can be reproduced by a Gaussian function. Moreover, $T_1$ is of single component, which also indicates that Ir is randomly distributed. For NMR Knight shift measurements, two single crystals, sized 2 mm - 4 mm - 0.5 mm and 2 mm - 3 mm - 0.4 mm, were used. For NQR measurements, the crystals were moderately crushed into grains to allow RF pulses to penetrate easily into the samples.  Small and thin single-crystal platelets cleaved from an as-grown ingot were used for X ray and dc-susceptibility measurements. 
 
\noindent
\textbf{NQR measurement.}

  For NQR, the nuclear spin Hamiltonian can be expressed as, $\mathcal{H}_{\rm Q}$ = $(h \nu_{\rm Q}/6)[3{I_z}^2-I(I+1)+\eta({I_x}^2-{I_y}^2)]$, where $h$ is Planck's constant and $I$ = 9/2 for the In nucleus is the nuclear spin; $\nu_{\rm Q}$ and the asymmetry parameter $\eta$ are defined as $\nu_{\rm Q}$  = $\frac{3eQV_{zz}}{2I(2I-1)h}$  and, $\eta$ $=$ $\frac{V_{xx} - V_{yy}}{V_{zz}}$, respectively, and $Q$ and $V_{\alpha\beta}$ are the nuclear quadrupole moment and EFG tensor, respectively. In CeRh$_{0.5}$Ir$_{0.5}$In$_5$, there are two inequivalent In sites, one in the CeIn [In(1)] layer and another in the (Rh,Ir)In$_4$ [In(2)] layer (see Fig. 2a). The  principle axis of the EFG at the In(1) [In(2)] site is parallel [perpendicular] to the $c$ axis. The $^{115}$In-NQR spectra for In(1) $\pm$1/2 $\leftrightarrow$ $\pm$3/2 transition line (Supplementary Note 2 and Supplementary Figure 2), $\pm$7/2 $\leftrightarrow$ $\pm$9/2 transition line (Supplementary Figure 5), the In(2) $\pm$3/2 $\leftrightarrow$ $\pm$5/2 transition line and $T_1$ for the In(1) site ($\eta = 0$)  (Supplementary Note 4 and Supplementary Figures 6 and 7) were obtained as reported in an earlier study \cite{Yamaguchi,recovery}.  Here, $T_1$ probes the dynamic spin susceptibility through the hyperfine coupling constant $A_{\bf q}$ as $1/T_1 \propto T\sum_{\bf q} \left | A_{\bf q} \right |^2 \chi ^{\prime\prime}({\bf q}, \omega_0)/\omega_0$,  where $\omega_0$  is the NQR frequency \cite{MoriyaJPSJ} and ${\bf q}$ is a wave vector for antiferromagnetic order and/or quantum critical fluctuation in CeRh$_{0.5}$Ir$_{0.5}$In$_5$.   Meanwhile, $1/T_1$ $\propto$ $N(E_{\rm F})^2$$k_{\rm B}T$ holds in a Pauli paramagnetic metal, i.e., in a heavy Fermi liquid state (Korringa law). Here, $N(E_{\rm F})$ is the density of states at $E_{\rm F}$.

  A NiCrAl/BeCu piston-cylinder-type pressure-cell filled with Daphne (7474) oil was used. The $T_{\rm c}$ of Sn was measured to determine the pressure. Measurements below 1 K were performed using a $^3$He refrigerator.   
  
\noindent
\textbf{X-ray measurement.}

   A diamond anvil cell filled with a CeRh$_{0.5}$Ir$_{0.5}$In$_5$ single crystal and Daphne oil were used for room temperature X-ray measurements under pressure; the pressure was determined by measuring the fluorescence of ruby. All measurements were made at zero magnetic field. 

\noindent
\textbf{EFG calculation.}

 The EFG is calculated using the all electron full-potential linear augmented plane wave method implemented in the Hiroshima Linear-Augmented-Plane-Wave (HiLAPW) code with generalized gradient approximation including spin-orbit coupling \cite{Oguchi}.

\noindent
\textbf{Hyperfine coupling constant.}

To estimate the hyperfine coupling constant for CeRh$_{0.5}$Ir$_{0.5}$In$_5$, $^{115}A(1)$, we measure the $^{115}$In(1)-NMR spectrum and dc susceptibility.  For NMR, the nuclear spin Hamiltonian is expressed as $\mathcal{H}$ = $-^{115}$$\gamma\hbar\bf{I}\cdot\bf{H}$$(1+K) + \mathcal{H}_{\rm Q}$, where  the gyromagnetic ratio $^{115}\gamma$ = 9.3295 MHz T$^{-1}$, $\bf{H}$ is the external magnetic field, and $K$ is the Knight shift. $^{115}$In-NMR spectra have nine transitions from $I_{\rm z}$ = (2$m$$+$1)/2 to (2$m$$-$1)/2 where $m$ = $-$4, $-$3, $-$2, $-$1, 0, 1, 2, 3, 4 for In(1) and In(2) sites, respectively, with $K$, $\nu_{\rm Q}$, and $\eta$ as parameters. At ambient pressure, $\nu_{\rm Q}$ and $\eta$ at the In(1) [In(2)] site are 6.35 MHz (17.37 MHz) and 0 (0.473), respectively \cite{Yamaguchi}. The Knight shift for In(1), $^{115}K(1)_{\rm c}(T)$, was calculated from the peak in the $^{115}$In(1)-NMR center line ($m$ = 0) taken by sweeping the RF frequency at a fixed field $H$ = 12.950 T and  $\chi_{\rm c}$ (emu mol$^{-1}$) is obtained by dc susceptibility measurements at $H$ = 3 T using a superconducting quantum interference device (SQUID) with the vibrating sample magnetometer (VSM).  The magnetic field $H$ is fixed perpendicular to the CeIn layer ($c$ axis).


\section*{Acknowledgments}
We would like to thank T. Kambe for help with the susceptibility measurement, Y. Fuseya, Y. Kuramoto,  H. Kusunose, J. Otsuki,  K. Miyake, S. Watanabe, T. Oguchi and Q. Si for discussion. This work was supported in part by research grants from MEXT (No. JP19K03747, JP23102717, and JP25400374), NSFC grant No. 11634015 and MOST Grant (No. 2016YFA0300300, No. 2017YFA0302904, and No. 2017YFA0303103) and, at Los Alamos, was performed under the auspices of the U.S. Department of Energy, Office of Basic Energy Sciences, Division of Materials Sciences and Engineering. 

 \section*{Authors contributions}
  G.-q.Z planned the project. J.L.S and J.D.T synthesized the single crystal. S.K, T.O, A.S, Y.K, and K.U performed NQR measurements. S.K and Y.K performed NMR and susceptibility measurements.  J.G, S.C, and L.L.S performed X-ray measurement. K.M performed band structure calculation. G.-q.Z, S.K and J.D.T wrote the manuscript.  All authors discussed the results and interpretation.

\end{document}